# Identifying and removing the cell-cycle effect from single-cell RNA-Sequencing data

Martin Barron, Jun Li[*]

Department of Applied and Computational Mathematics and Statistics, University of Notre Dame, Notre Dame, IN 46556, USA

[*]To whom correspondence should be addressed (email: jun.li@nd.edu).

**Abstract**

Single-cell RNA-Sequencing (scRNA-Seq) is a revolutionary technique for discovering and describing cell types in heterogeneous tissues, yet its measurement of expression often suffers from large systematic bias. A major source of this bias is the cell cycle, which introduces large within-cell-type heterogeneity that can obscure the differences in expression between cell types. The current method for removing the cell-cycle effect is unable to effectively identify this effect and has a high risk of removing other biological components of interest, compromising downstream analysis. We present ccRemover, a new method that reliably identifies the cell-cycle effect and removes it. ccRemover preserves other biological signals of interest in the data and thus can serve as an important pre-processing step for many scRNA-Seq data analyses. The effectiveness of ccRemover is demonstrated using simulation data and three real scRNA-Seq datasets, where it boosts the performance of existing clustering algorithms in distinguishing between cell types.

Identifying and characterizing different cell types in heterogeneous tissues is the foundation of understanding how cancer evolves and metastasizes, how brains function, how stem cells program and develop, among numerous other important applications. However, this cannot be done using the regular (bulk-based) RNA-Sequencing technique, which is the *de facto* standard for measuring the transcriptome but can only measure the average expression of all cells in bulk. ScRNA-Seq eliminates these limitations by preparing libraries from single cells and measuring the individual transcriptional profiles of hundreds or thousands of single cells (See e.g. [1–8] for reviews).

Applying clustering algorithms, such as k-means clustering or hierarchical clustering, to the gene expression profiles of single cells can reveal the different cell types present in heterogeneous tissues, allowing them to be identified and characterized[9–14]. However, for this approach to achieve its optimum power the high-noise nature of scRNA-Seq data needs to be carefully handled[15–21]. ScRNA-Seq data, while known to have large variance introduced during library preparation[17,22], also suffers from large systematic bias caused by biological noises, which act as confounding factors that obscure biological signals of interest in the data[12,15,23]. For data generated by other high-throughput techniques such as microarrays, removing systematic bias has been shown to be critically important[24–26]. For scRNA-Seq data, one of the major sources of biological noise is the cell cycle[19,27–32]. During the cell cycle a cell increases in size, replicates its DNA and splits into daughter cells. Different cells are at different time points of this cycle, and thus they may have quite different expression profiles[15], even if they are cells of the same type[33,34]. This within-type heterogeneity can seriously deteriorate the performance of clustering algorithms for cell type identification: it may blur clusters of cell types or cause cells of similar cell-cycle statuses to stand out as new clusters. Fig. 1 shows an example using simulation data. Gene expression data is simulated for 50 cells and 2,000 genes. The cells are randomly assigned to two cell types (denoted using different shapes) and three cell-cycle stages (denoted using different colors). Fig. 1a shows the results of principal component analysis (PCA) on this simulated data. The cells are clustered into six distinct clusters, grouping by both cell types and cell-cycle statuses. Cell-type discovery using this original data directly will mistakenly result in the discovery of six cell types.

The aim of this paper, is to develop an efficient computational method to remove this effect from the data, giving a dataset free from the cell-cycle effect, on which downstream analysis, such as discovering cell types, can be more efficient.

Some genes, from annotation databases, are known to play a role in the cell cycle and their expressions are heavily influenced by the cell cycle. These genes are often called "cell-cycle genes."[12,35] However, attempting to remove the cell-cycle effect by simply excluding these cell-cycle genes from the analysis is not a viable strategy. This is because the cell cycle also affects the expression level of many genes which are thought to be unrelated to the cell cycle[12], although usually to a lesser extent compared to the cell-cycle genes. For example, when considering a set of over 6,500 genes not previously associated with the cell cycle, Buettner et al.[12] found that 44% of the genes showed significant correlation with at least one cell-cycle gene.

The scLVM (single-cell latent variable model) algorithm first proposed the idea of estimating the cell-cycle effect and then removing this effect from scRNA-Seq data[12]. All genes are retained after applying scLVM, but the effect of the cell cycle will be removed from their expression levels. scLVM uses only the cell-cycle genes to identify the cell-cycle effect. It develops a sophisticated Bayesian latent variable model to reconstruct hidden factors in the expression profile of the cell-cycle genes. It declares that the leading $K$ ($K \geq 1$) factors are the cell-cycle effect and removes them from the whole dataset. No formal statistical methods have been proposed to choose $K$ with the authors recommending using either the default value $K = 1$, or relying on a scree plot of the variance captured by each latent factor and using the elbow point, similar to choosing the number of significant components in a PCA. scLVM has shown its ability in removing the cell-cycle effect from a real scRNA-Seq dataset, which is the first real data example we will show in the Results section. To date, scLVM is still the only available method for removing the cell-cycle effect. The key assumption that scLVM makes is that all the main effects in the expression of cell-cycle genes are cell-cycle effects. However, we have realized that this may not hold, making the application of scLVM hazardous. Cells of different cell types, even if they are in the same time point of their cell cycle, should have different expressions even on cell-cycle genes. In other words, the expression of cell-cycle genes is also influenced by the cell type. We call the expression change caused by the cell type "the cell-type effect." Likewise, there can be effects from experimental condition, disease state, etc. There is no guarantee that the cell-cycle effect is stronger than all other effects, even on the cell-cycle genes. Indeed even if the cell-cycle effect is the strongest, it is not likely that all its components (generally, more than one latent factor is needed to describe the cell-cycle effect) are stronger than the components of other effects. In other words, some of the leading $K$ factors of the gene expression profile of the cell-cycle genes may not be generated by the cell cycle and instead may originate from biological features of interest such as differences in cell type. Removing all the leading $K$ factors will remove these signals of interest from the data, compromising the downstream analysis of the data, such as clustering analysis for cell-type discovery, defeating the purpose of a scRNA-Seq experiment. For clearer illustration, we show four cases in Table 1 as examples. In case 1, the first leading factor in the cell-cycle genes represents the cell-cycle effect; scLVM will work when $K = 1$ is used. In case 2, the top two leading factors in the cell-cycle genes both represent the cell-cycle effect; scLVM will work when $K = 2$ is used, although the cell-cycle effect will not be removed completely when the default value $K = 1$ is used and other effect(s) may be removed along with the cell-cycle effect if $K > 2$ is used. In case 3, the first leading factor represents another effect of interest; scLVM will remove this effect no matter what $K$ value is used, meaning that scLVM will always fail. In case 4, the first and third leading factors represent the cell-cycle effect; scLVM will not remove the cell-cycle effect completely (when $K \leq 3$) and/or remove other effects as well (when $K \geq 2$).

A better method should include a mechanism to check each factor and make a judgement as to whether the factor represents the cell-cycle effect. We propose a method called c̲ell-c̲ycle remover (ccRemover) that effectively identifies the components of the cell-cycle effect from scRNA-Seq data. It then removes them from the data while preserving the other components of the data. ccRemover identifies the cell-cycle effect using the expression profiles of all genes. For simplicity, we call genes that are not annotated as cell-cycle genes "control genes." The assumption that ccRemover makes is that the cell-cycle effect is stronger on average in the cell-cycle genes than control genes.

ccRemover carries out a simple PCA on the expression profiles of control genes to capture the sources of variation/effects, represented by the loadings of the principal components. It then projects the expression of all genes on these loadings to get the component scores for each gene. The magnitude of the component scores measures the strength of the effects on the genes. For each effect (principal component), ccRemover compares the average magnitude of the component scores of the cell-cycle genes with the average magnitude of the component scores of control genes. It declares all effects whose average magnitude is larger on the cell-cycle genes than on the control genes as cell-cycle effects. A formal bootstrap-based statistical test is developed for this comparison. Then all effects declared as the cell-cycle effect are removed from the whole dataset by subtracting the projections of gene expression profiles on these effects. This identification and removal process is repeated until no more principal components are identified as the cell-cycle effect.

The only assumption that ccRemover makes is that the cell-cycle effect is stronger in the cell-cycle genes than control genes, "on average." It does not matter if some annotated cell-cycle genes are not truly influenced by the cell cycle, or if some control genes are directly involved into the cell-cycle process. As long as the set of cell-cycle genes are better than random picks from all genes, the assumption of ccRemover holds. Thus, ccRemover is insensitive to the completion of annotation databases in assigning cell-cycle genes.

The performance of ccRemover is demonstrated using a simulated dataset and three real scRNA-Seq datasets, where ccRemover is able to successfully remove the effects of the cell cycle from the data while preserving the other components of the data. We show that ccRemover can aid in the identification of subpopulations of cells, improve the clustering analysis of single cells and performs favorably compared to scLVM.

# Results

For the simulated data and each of the real scRNA-Seq datasets the analysis follows the same process. Firstly we apply scLVM and ccRemover to the original dataset, which we call "the original data." scLVM is applied using the python script available from the scLVM GitHub page. This gives us a scLVM corrected dataset and a ccRemover corrected dataset, which we refer to as the "scLVM corrected data" and the "ccRemover corrected data" respectively. Once we have acquired the three datasets the same clustering algorithms and statistical tests are applied to each of them allowing us to compare the performance of the methods.

We use the same set of cell-cycle genes when applying scLVM and ccRemover. The lists of cell-cycle genes are acquired by combining two sources. Firstly Biomart was used to download lists of genes that were annotated to the cell cycle[38]. In addition two R packages were used to retrieve gene annotation data from GO term[39], and these were *org.Mm.eg.db*[40] and *org.Hs.eg.db*[41] for annotations for mouse and human respectively. For the choice of $K$ (the number of leading factors to be removed) in scLVM, we try both the default value $K = 1$ and the value given by the scree plot.

## Simulation Data

We simulate data matrix $\boldsymbol{X}$ that contains measurements for 50 cells and 2,000 genes, of these genes 400 are assigned as cell-cycle genes. The cells are randomly assigned to the two classes (cell types) and three cell-cycle stages. Suppose cell $j$ is assigned to class $t_j$ and cell-cycle stage $s_j$, $t_j \in \{1, 2\}$ and $s_j \in \{1, 2, 3\}$. We simulate $X_{ij}$, the expression of gene $i$ in cell $j$ by

$$X_{ij} = Y_{it_j} + Z_{is_j} + W_{ij},$$

where $Y$ is the cell-type effect, $Z$ is the cell-cycle effect, and $W$ is random noise. The cell-type effect is generated by $Y_{i1}, Y_{i2} \sim N(0, 1.2^2)$, the cell-cycle effect is generated by $Z_{i1}, Z_{i2}, Z_{i3} \sim N(0, 1)$ for cell-cycle genes and $Z_{i1}, Z_{i2}, Z_{i3} \sim N(0, 0.6^2)$ for control genes, and the random noise is generated by $W_{ij} \sim N(0, 1)$.

In Fig. 1, the data is plotted on the first two principal components with the shape of the points corresponding to their cell type and the color corresponding to their cell-cycle stage. Ideally, the data should be separated only by shape and not color, that is, by cell type and not cell-cycle stage. However, on the original data (Fig. 1a), the cells are clustered into six different groups corresponding to the cell type and cell-cycle stage combinations demonstrating how the cell cycle can confound the analysis of scRNA-Seq data.

scLVM removes the first leading factor ($K = 1$, default choice) or the first three leading factors ($K = 3$, suggested by the scree plot). Fig. 1b shows the results when the first leading factor is removed, where the cells are clustered into three groups according to the cell-cycle stage, and cells from different cell types are completely indistinguishable. scLVM has failed completely here by mistakenly removing the cell-type effect instead of the cell-cycle effect. Fig. 1c shows the results when all three leading factors are removed. The cells exhibit no clear clusters, indicating that scLVM has removed both the cell-cycle effect and the cell-type effect. The data has effectively been rendered useless as it now contains just noise.

Fig. 1d shows the results of correcting the data using ccRemover, where the cells are well separated by the cell type and within each cluster cells with different cell-cycle stages are completely mixed. This means that the cell-cycle effect has been thoroughly removed, while the cell-type information has been preserved. ccRemover is able to correctly identify the second and third principal components as cell-cycle effects and removes them.

In our simulation study above, we made two simplifications in the data simulation. First, we simulated Gaussian data directly instead of simulating the raw count data, normalizing the counts by the sequencing depth, and then log-transforming them. Second, we simulate the cell cycle as three discrete stages. In reality, the cell-cycle status is more like a continuous variable, as even if two cells are in the same stage, they may still differ in how far they have progressed through that stage.

In our simulation, the leading latent factor of the expression profile of cell-cycle genes is not the cell-cycle effect. This corresponds to case 3 in Table 1, and scLVM fails as expected. Changing our simulation parameters can make the data represent other cases in Table 1, on many of which we should not expect such distinct results between scLVM and ccRemover. We will show a wider range of comparisons using real datasets.

## Real dataset 1: T helper cell data

The first real dataset is the differentiating T-helper ($T_H$) cell dataset that was used to display the ability of scLVM to help reveal hidden subpopulations of cells by Buettner et al.[12]. We will demonstrate that ccRemover also has this ability, and improves on the performance of scLVM. The dataset was generated by Mahata et al.[42] to study the differentiation of $T_H$ cells and the steroids they synthesize to contribute to immune homeostasis. The data was created by polarizing naive $T_H$ cells *in vitro* towards a $T_H2$ subtype, leading to a population in which there are cells differentiating towards the $T_H2$ subtype and cells which are not. The objective for this dataset is to identify biologically meaningful clusters of cells. The original dataset was downloaded from the supplementary materials of Buettner et al.[12] and contains normalized and log transformed expression measurements for 81 cells and 7,073 genes, of which 532 were identified as cell-cycle genes. For this dataset, the scLVM corrected data along with cluster assignments for the corrected data are also available from the same source and were used to evaluate the performance of scLVM. When ccRemover is applied to the original data it identifies the first principal component to be a cell-cycle effect on the first iteration. Once this effect is removed from the data no other features are deemed to be cell-cycle related.

Both scLVM and ccRemover remove the cell-cycle effect efficiently on this data. To check this, in Fig. 2, we plot the density of the expression level of cell-cycle genes selected from the top ranked genes on Cyclebase[43]. On the original data (red lines), many genes display a bimodal density commonly seen in scRNA-Seq data indicating the on-off action of genes, in this case, controlled by the cell cycle. On the scLVM (green lines) or ccRemover (blue lines) corrected data, the bimodality of the densities largely disappears and most genes display a unimodal distribution indicating that the cell-cycle effect has been reduced or removed completely for these genes.

To determine if biologically meaningful clusters can be discovered from the data we avail of a criterion for measuring performance used by Buettner et al. during their analysis. There is a list of 122 known $T_H2$ signature genes curated by Buettner et al. If the cells are clustered into two clusters and genes that are differentially expressed between these two clusters are identified, these $T_H2$ signature genes should be over-represented in the set of differentially expressed genes if different clusters represent physiologically distinct subpopulations of cells. This over-representation can be summarized by an odds ratio of the percentage of $T_H2$ signature genes in the set of differentially expressed genes to that in all genes. A large odds ratio is favored.

To implement this criterion, we applied 2-means clustering and use a t-test with false discovery rate 0.01 to identify differentially expressed genes. Then the odds ratio, the 95% confidence interval of the odds ratio, and the p-value of the hypothesis of $\text{odds ratio} > 1$ were calculated by a hypergeometric test. The results are shown in Table 2. On the original data, the odds ratio is less than 1, indicating that the clustering of cells is unlikely to be physiologically meaningful. The true substructure of the data is completely obscured, and this could be due to the confounding effects of the cell cycle.

On the scLVM corrected data, the odds ratio is 2.382, with the lower confidence interval bound of 1.518 and p-value $1.542 \times 10^{-4}$. On the ccRemover corrected data, the odds ratio is 3.439, with the lower confidence interval bound 2.297 and p-value $2.373 \times 10^{-9}$. This indicates that both scLVM and ccRemover are able to remove the cell-cycle effect from the data so that the true substructure of the data can be revealed, and ccRemover removes the cell-cycle effect more thoroughly and/or keeps other biological features more intact compared to scLVM.

## Real dataset 2: human glioblastomas data

This dataset contains cells from five human glioblastomas[44]. It was created by Patel et al. by isolating individual cells from freshly resected and dissociated IDH1/2 wild-type primary human glioblastomas, MGH26, MGH28, MGH29, MGH30 and MGH31. This dataset has log transformed and centered TPM (Transcripts Per Million) measurements for 5,948 genes and 430 single cells with each tumor represented by 70 to 118 cells. Of the 5,948 genes 412 were identified as cell-cycle genes. It has been shown that the level of cell-cycle activity within this dataset is very low, with an average of only 8% of the cells per tumor showing cell-cycle activity[44]. For this dataset the objective is to cluster the cells by their tumor of origin.

When scLVM was applied, the scree plot suggests removing the first leading hidden factor, agreeing with its default choice. When ccRemover was applied to this dataset the $5^{th}$, $6^{th}$ and $9^{th}$ components were identified as cell-cycle effects and removed on the first iteration. On the second iteration the $10^{th}$ component was identified as a cell-cycle effect. Once this effect was removed from the data there were no more cell-cycle effects detected.

Hierarchical clustering was applied to the (original, scLVM corrected, and ccRemover corrected) data, splitting the cells into five clusters, with each cluster being assigned the class of the majority of the cells contained within the cluster. The results are shown in Fig. 3. On the original data, 87.44% of the cells were clustered correctly. From the plot of the dendrogram (Fig. 3a) it is clear that the MGH31 (red) cluster contains cells from all the other tumors that have been incorrectly classified, the MGH28 (purple) and MGH30 (blue) clusters also display significant impurities. On the scLVM corrected data, 90.00% of the cells were classified correctly, an improvement of over 2.5% from the original data. On the ccRemover corrected data, 92.32% of the cells were classified correctly, an increase of nearly 5% from the original data. The purity of the clusters in the dendrogram (Fig. 3c) for the ccRemover corrected data show marked improvement over the original data, and especially the MGH28 and MGH31 clusters show convincing improvements in purity. This result is particularly striking when considering the very low levels of cell-cycle activity within this dataset and demonstrates that ccRemover can improve the downstream analysis of scRNA-Seq data even when the cell-cycle effect is not very strong.

**Real dataset 3: lung adenocarcinoma data**

This dataset was generated by Kim et al. to investigate the mechanisms by which intra-tumoral heterogeneity impacts the therapeutic outcome of cancer treatments[45]. It contains 176 cells from three cell types: 77 patient-derived xenograft tumor cells from a lung adenocarcinoma patient tumor xenograft, 50 single H358 human lung cancer cells (H358), and 49 PDX cells derived from a lung cancer-brain metastasis (LC.MBT). Interestingly, the 77 cells in the first type come from two groups of cells that are technical replicates of each other. One group contains 34 cells, and the other group contains 43 cells, and they are called LC.PT and LC.PT_RE in the original paper. These two groups of cells were isolated and RNA-sequenced separately, and thus there should be batch effects, which may affect specific subsets of genes and may affect different genes in different ways[46].

The TPM values for 57,820 genes are available for each of the 176 cells. Prior to analysis any genes which had zero expression for over two thirds of the cells were removed from the data, leaving 10,977 genes of which 757 were annotated to the cell cycle. The data was transformed to a log-scale by adding 1 to each of the measurements and taking the natural log.

The scree plot from scLVM suggests removing the first leading hidden factor, agreeing with its default choice. Instead, ccRemover suggests removing six principal components in its four iterations, and interestingly, these six components do not include the very first principal component.

When using 3-means clustering on the original data, the three clusters represent the three cell types perfectly, and thus there is no room for improvement. Instead, we consider using 4-means clustering, in order to see whether the 77 cells of the first cell type can be clustered accordingly to the two sets of technical replicates, LC.PT and LC.PT_RE. Fig. 4 shows the results. On both the original data (Fig. 4a) and the scLVM corrected data (Fig. 4b), the LC.PT and LC.PT_RE cells are split into two clusters (clusters 3 and 4) each containing roughly equal proportions of cells from each set, indicating that the technical replicates are non-separable. On the ccRemover corrected data (Fig. 4c), on the other hand, the majority (80%) of cluster 3 are cells from the LC.PT_RE group, while the majority (89%) of cluster 4 are cells from the LC.PT group. This means that cells from different sets of technical replicates are largely separated by the batch effect. This batch effect is present in all three of the original and corrected datasets, but it has a noticeable influence in the clustering results only on ccRemover corrected data. The reason could be that the batch effect is confounded by the stronger cell-cycle effect in the original data, and it stands out when the cell-cycle effect was removed by ccRemover. scLVM may have not removed the cell-cycle effect thoroughly enough to make a difference.

Further analysis was carried out to determine if this is the case. Fig. 5 displays heat maps of the expression of the top ranked cell-cycle genes from Cyclebase[43]. The cell-cycle genes displayed in the heat map are ordered based on the time point of the cell cycle at which their expression peaks. If the cell-cycle effect exists, there should be blocks of similar expression levels, and these blocks should not occupy from the first row to the last row as the genes do not achieve their peak expressions at the same time point of the cell cycle. On the original data (Fig. 5a), there are clear such blocks, and the most prominent one is shown in a blue box. For the scLVM corrected data the blocks are less apparent but still present (Fig. 5b), indicating that the cell-cycle effect has been removed partially. For the ccRemover corrected data (Fig. 5c), there are no easily visible blocks left indicating that ccRemover has effectively removed the cell-cycle effect from this dataset. For both the scLVM and ccRemover corrected data the range of expression for the cell-cycle genes is reduced and so the heat map colors show less variation.

This example shows a feature of ccRemover: while it quite thoroughly removes the cell-cycle effect, it keeps all other effects, favorable (like the cell-type effect) or unfavorable (like the batch effect), intact. This is exactly what ccRemover is designed for. In this example, ccRemover makes the batch effect stand out, which may actually facilitate removing the batch effect. This can be done by using software specifically designed for removing the batch effect, and it is out of scope of this paper.

## Discussion

ScRNA-Seq data suffer from a systematic bias which is introduced by the cell cycle. The cell cycle can have a confounding effect on the analysis of scRNA-Seq data, conceal the true biological features of interest and compromise the interpretation of scRNA-Seq experiments. In fact, we saw for the differentiating T-cell data that the true substructure of the data was undetectable unless the cell-cycle effect was removed. This can increase the difficulty of identifying new subtypes and subpopulations of cells in scRNA-Seq data. The current method developed for removing this effect, scLVM, does not inspect whether a leading factor represents the cell cycle, and thus it has a considerable risk of removing other important features of the data, as well as removing the cell-cycle effect incompletely. We developed a new method, ccRemover, that includes a formal statistical test to inspect whether an effect is a component of the cell-cycle effect or not. By using this test, ccRemover is able to remove the effects of the cell cycle from scRNA-Seq data quite thoroughly while preserving the other information that is contained within the data. Applying ccRemover to remove the cell-cycle effect can allow previously distorted signals of interest to emerge from the data and improve the analysis of scRNA-Seq data. This has been shown in both simulation data and three real datasets.

The cell cycle is often the main source of biological noise in scRNA-Seq data. When it is removed by ccRemover, other effects may stand out as the main confounding factors, as we have shown in our third real data example. ccRemover does not remove these effects as it is designed for removing the cell-cycle effect exclusively. However, if a set of genes are known to be more influenced by a particular effect to be removed, one can treat this set of genes as set $A$ (the cell-cycle genes) and then ccRemover can be directly used to remove this effect.

## Methods

We describe our ccRemover algorithm in this section. Denote the matrix of gene expression values as $X$, with element $X_{i,j}$ being the expression value for the $i^{th}$ gene and the $j^{th}$ cell, $i = 1, \ldots, p$ and $j = 1, \ldots, n$. We recommend transforming the data to a log scale and centering it on a gene-by-gene basis. Let $A = \{i : \text{gene i is a cell-cycle gene}\}$ and $B = \{i : \text{gene i is a control gene}\}$, with the corresponding data matrices $X_A$ and $X_B$. The numbers of genes in $A$ and $B$ are represented as $|A|$ and $|B|$ respectively. Thus, the dimensions of $X$, $X_A$ and $X_B$ are $n \times p$, $|A| \times p$, and $|B| \times p$, respectively.

The ccRemover algorithm follows these steps:

1. Perform a PCA on the data matrix of control genes $X_B$. Let the loadings be $v_1, \ldots, v_n$, and the corresponding component scores be $\beta_1, \ldots, \beta_n$ with $\beta_j = X_B v_j$. Then $X_B = \sum_{j=1}^{n} \beta_j v_j^T$.

2. Project the data matrix of cell-cycle genes $X_A$ onto $v_1, \ldots, v_n$. The component scores for $v_j$ are
$$\alpha_j = X_A v_j.$$

3. Find the set of $v_1, \ldots, v_n$ that have significantly larger component scores on cell-cycle genes than on control genes. This can be done by testing whether

$$\Delta_j = \frac{\|\alpha_j\|_2}{\sqrt{|A|}} - \frac{\|\beta_j\|_2}{\sqrt{|B|}} > 0.$$

   using the bootstrap (details given later), where $\| \cdot \|_2$ denotes the L2 norm. Let the significant set of $j$ be $S$. The directions $\{v_j; j \in S\}$ will be used as the cell-cycle effect.

4. Project $X$ onto $v_j$, $j \in S$ to extract the cell-cycle effect from the data matrix. Subtract these projections from $X$ to remove them from the data. That is, the corrected data matrix with the cell-cycle effect removed is given by $\hat{X} = X - \sum_{j \in S}(X v_j) v_j^T$.

Steps 1 to 4 are repeated until no more cell-cycle effects are identified (i.e. no statistically significant $\Delta_j > 0$ is found). We have found that usually no more than four repetitions are needed.

We use the following two-class bootstrap procedure[47] to test whether $\Delta_j$ is significantly larger than 0:

1. Take a random sample with replacement of $|A|$ columns from $X_A$ and another $|B|$ columns from $X_B$. This gives the resampled data matrices $X_A^*$ and $X_B^*$.

2. Calculate $\Delta_j^*$, a bootstrap replicate of $\Delta_j$, by applying steps 1 and 2 of the algorithm of ccRemover to $X_B^*$ and $X_A^*$.

3. Repeat steps 1 and 2 of this algorithm $n_{\text{boot}}$ times to get bootstrap replicates $\Delta_j^{*1}, \ldots, \Delta_j^{*n_{\text{boot}}}$. We use $n_{\text{boot}} = 100$ for all our simulations and real data examples. Let the standard deviation of these bootstrap replicates be $\sigma_{\Delta_j}$.

4. Reject $H_0 : \Delta_j \leq 0$ when the bootstrap based t-statistic $\frac{\Delta_j}{\sigma_{\Delta_j}} \geq C_\Delta$, where $C_\Delta$ is a cutoff specified by the practitioner.

For most datasets, we suggest using $C_\Delta = 3$, which roughly corresponds to a p-value of 0.01. We used this cutoff for all our simulations and real data examples except for the glioblastoma data, where it is known that the cell-cycle activity is at a very low level[44]. We use a smaller cutoff value $C_\Delta = 2$ for this data, and it roughly corresponds to a p-value of 0.05.

# Tables

Table 1 – The performance of scLVM depends on the type of effect each leading factor describes

|  | Case 1 | Case 2 | Case 3 | Case 4 |
|---|---|---|---|---|
| Leading factor #1 | Cell-cycle | Cell-cycle | Other | Cell-cycle |
| Leading factor #2 | Other | Cell-cycle | Cell-cycle | Other |
| Leading factor #3 | Other | Other | Cell-cycle | Cell-cycle |
| Leading factor #4 | Other | Other | Other | Other |
| Performance of scLVM | Likely good | Maybe good | Fail | Poor |

Table 2 – Statistical tests on the differentiating T-cell dataset.

| Method | odds ratio | 95% confidence interval of the odds ratio | p-value |
|---|---|---|---|
| original data | 0.466 | (0.318, 0.687) | 0.999 |
| scLVM | 2.382 | (1.518, 3.655) | $1.542 \times 10^{-4}$ |
| ccRemover | 3.439 | (2.297, 5.100) | $2.373 \times 10^{-9}$ |

# Figures

Figure 1 – The simulation data projected onto its first two principal components. The cell types are represented by the different shapes (circle, triangle) and the cell-cycle time point of each cell is represented by the different colors (red, blue, green). **(a)** Original Data. Here the data is clustered into six groups corresponding to the combinations of cell type and cell-cycle status. **(b)** scLVM corrected data (one latent factor removed). The data clusters into three groups corresponding to cell-cycle status. **(c)** scLVM corrected data (three latent factors removed). No distinct clusters are observed. **(d)** ccRemover corrected data. The data splits into two groups corresponding to the cell types.

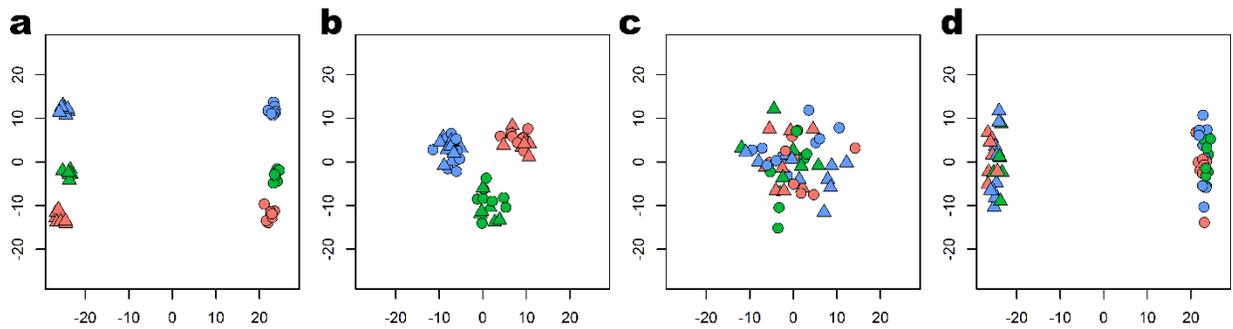

Figure 2 - Density plots of selected genes from the T-cell data. The densities are displayed for the original (red), scLVM corrected (green) and ccRemover corrected (blue) data. The genes were selected from among the top ranked genes on Cyclebase[43]. The original data displays bimodal densities which are common in scRNA-Seq data indicating genes whose expression switches on and off[18,48–50]. When the cell-cycle effect is removed using ccRemover or scLVM these bimodal densities disappear.

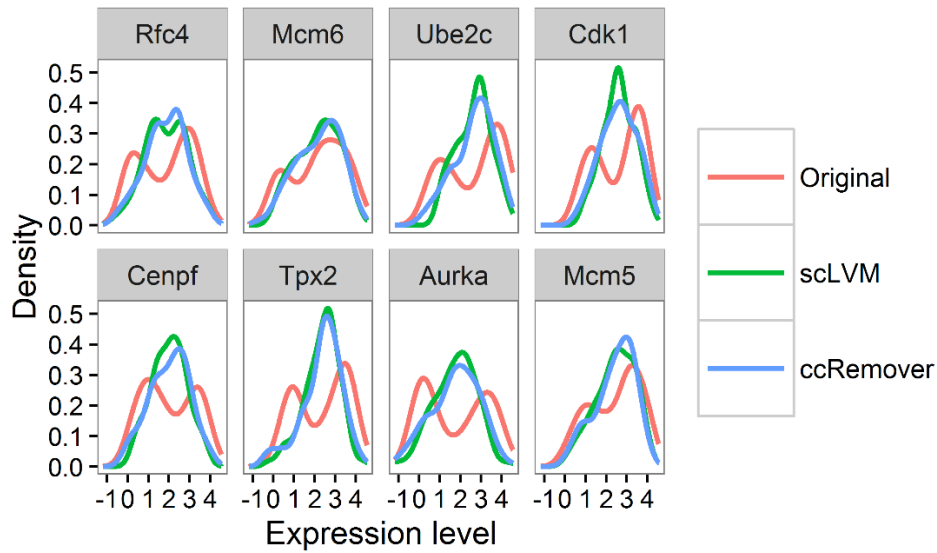

Figure 3 – Dendrogram plots from the hierarchical clustering on the original, ccRemover corrected and scLVM corrected glioblastoma data. The tumor of each of the cells is represented by their colors, MGH26 (yellow), MGH28 (purple), MGH29 (orange), MGH30 (blue) and MGH31 (red). The clustering assignments are displayed as boxes separating the cells. **(a)** Original data. There are significant misclassifications within the clusters for the original dataset. In particular the MGH28, MGH30 and MGH31 clusters contain significant numbers of cells from the other tumors. **(b)** scLVM corrected data. There is an increase in the accuracy of the clustering from the original data, however the MGH26 and MGH30 cells are now mixed between clusters. **(c)** ccRemover corrected data. There is a significant improvement in the purity clusters here compared to the original and scLVM corrected data. The MGH28 cluster is now much purer and only contains a few cells from the other tumors.

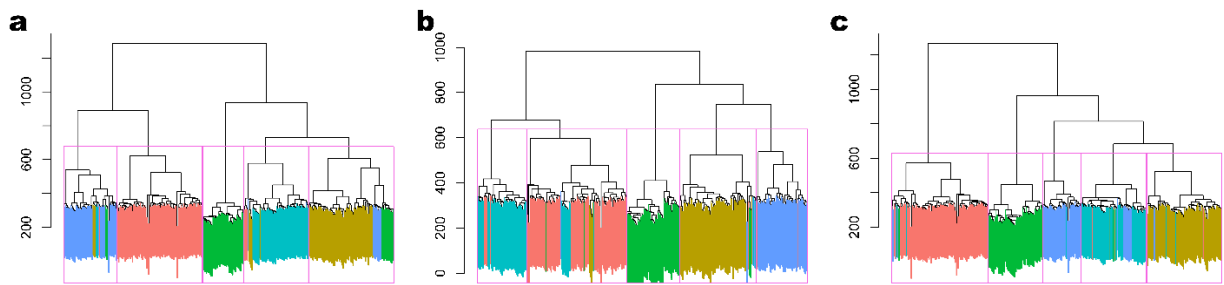

Figure 4 – Bar plots of the clustering assignments for the lung adenocarcinoma cells. **(a)** Original data. The LC.PT and LC.PT_RE cells split into two clusters each containing a roughly equal proportion of cells from each sample, indicating that 4-means failed to separate the cells from these two samples. **(b)** scLVM corrected data. Similar to the original data scLVM fails to split the LC.PT and LC.PT_RE cells into separate clusters. **(c)** ccRemover corrected data. The separation of the LC.PT and LC.PT_RE cells between the clusters has improved significantly with one cluster dominated by LC.PT cells and the other by LC.PT_RE cells.

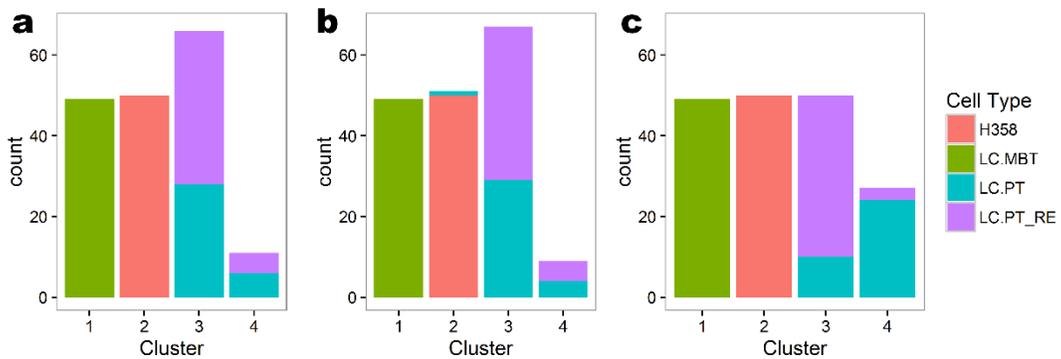

Figure 5 – Heat maps of gene expression in the lung adenocarcinoma dataset. The cell-cycle genes were chosen from the top ranked cell-cycle genes on Cyclebase[38] and are ordered by their cell-cycle peak time. The cells were ordered based on a hierarchical clustering of the original data and the order is the same for each heat map. **(a)** Original Data. The blocks of similar expression indicate cells at a similar cell-cycle time point, indicating the presence of cell-cycle effects. **(b)** scLVM corrected data. The blocks of similar expression have been reduced but are still apparent. The color of the heat map is more balanced as the range of the expression levels is reduced after they have been corrected. **(c)** ccRemover corrected data. The obvious blocks have been removed from the corrected dataset.

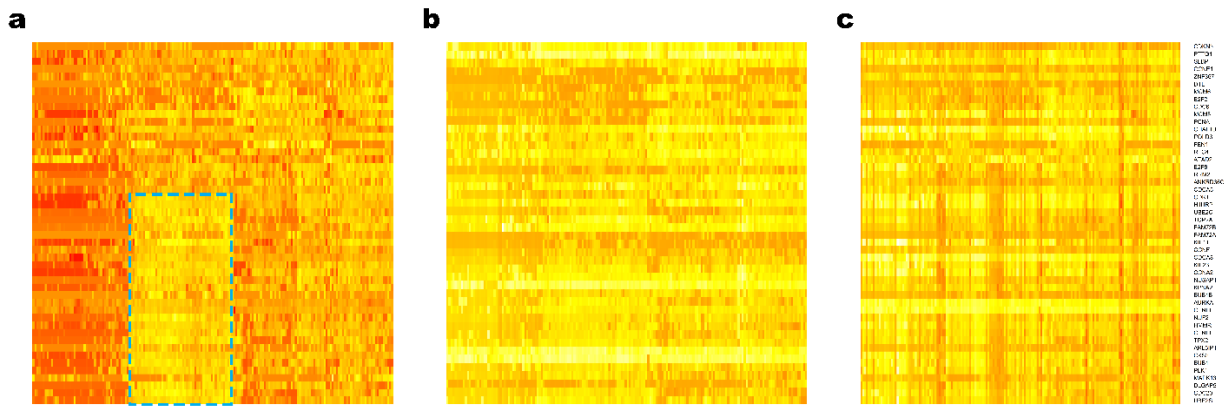